\documentclass{article}

\usepackage[preprint]{neurips_2026}
\usepackage{amssymb}

\usepackage[utf8]{inputenc} % allow utf-8 input
\usepackage[T1]{fontenc}    % use 8-bit T1 fonts
\usepackage{hyperref}       % hyperlinks
\usepackage{url}            % simple URL typesetting
\usepackage{booktabs}       % professional-quality tables
\usepackage{amsfonts}       % blackboard math symbols
\usepackage{nicefrac}       % compact symbols for 1/2, etc.
\usepackage{microtype}      % microtypography
\usepackage{xcolor}         % colors
\usepackage{amsmath}
\usepackage{graphicx}
\usepackage{svg}
\usepackage{gensymb}

\newcommand{\pos}[0]{\mathbf{r}}

\newcommand{\nin}[0]{\noindent{}}
\newcommand{\hasnt}[0]{$\times$}
\newcommand{\has}[0]{$\checkmark$}
\title{Expander attention as exchange-correlation}

\author{%
  Karim K. Alaa El-Din \\
  Department of Physics\\
  University of Oxford\\
  Clarendon Laboratory, Parks Road, Oxford OX1 3PU, UK \\
  \texttt{karim.alaael-din@physics.ox.ac.uk} \\
  \And
  Antonius v. Strachwitz \\
  Department of Physics\\
  University of Oxford\\
  Clarendon Laboratory, Parks Road, Oxford OX1 3PU, UK \\
  \And
  Sam M. Vinko\thanks{Also at Central Laser Facility, STFC Rutherford Appleton Laboratory, Didcot
OX11 0QX, UK.} \\
  Department of Physics\\
  University of Oxford\\
  Clarendon Laboratory, Parks Road, Oxford OX1 3PU, UK \\
  \texttt{karim.alaael-din@physics.ox.ac.uk} \\
}

\begin{document}

\maketitle

\begin{abstract}
Kohn-Sham density functional theory (DFT) is the workhorse of quantum chemistry, offering an attractive balance between accuracy and computational cost. Although exact in principle, DFT in practice relies on an approximation to the unknown exchange-correlation (XC) functional, which encodes the many-body quantum effects beyond the mean-field treatment. Many such approximations exist, and machine-learned XC functionals have proliferated in recent years. A persistent challenge in this area is the trade-off between accuracy and computational cost: while high-accuracy ML functionals have shown success on strongly correlated systems that are notoriously difficult for conventional approximations, their unfavorable scaling has limited broader adoption. Here, we propose a linearly scaling non-local XC approximation based on an expander graph transformer ansatz, improving the scaling of $O(N^2)$ or worse for previous ML functionals capable of reliably capturing strongly correlated systems. We show that it recovers the correct $\mathrm{H_2}$ dissociation curve in the strongly correlated regime, with promising results on planar $\mathrm{H_4}$, a system where even high-level coupled cluster methods break down. Our approach thus charts a path toward ML functionals that are both accurate on strongly correlated systems and cheap enough to deploy at scale.
\end{abstract}

\section{Introduction}
Density functional theory (DFT) is known as the workhorse of quantum chemistry due to its favourable combination of high computational efficiency and sufficient accuracy for most applications~\cite{Jones2015DensityFuture}. These properties have led to its usage in fields as diverse as drug discovery~\cite{Cavalli2006Target-RelatedDesign}, hydrogen storage technologies~\cite{Cruz-Martinez2024DensityMaterials}, next generation solar cells~\cite{Ganji2023CharacterizationCells}, and the study of planetary interiors~\cite{Stixrude2009ThermodynamicsEarth}. Kohn-Sham (KS) DFT in particular has proven a useful tool across scientific disciplines, providing a reliable means to get high-quality electronic structure calculations for both molecular systems and solids. At its core lies the abstraction from the treatment of individual electrons as quantum particles to an electronic density in space. This density is then found by self-consistently solving the Schrödinger-like KS equations. While this framework is in principle exact, errors enter through the so-called exchange-correlation (XC) functional, which embodies all effects neglected by this density treatment, including Pauli exchange for example. This term should be universal~\cite{Hohenberg1964InhomogeneousGas}, but it has been demonstrated that it would be Quantum Merlin Arthur hard to compute~\cite{Schuch2009ComputationalFunctionaltheory}. Instead, hundreds of density functional approximations (DFAs) have been proposed in its place~\cite{Lehtola2018}. This veritable zoo of DFAs is composed of a range of different approximations, each with their own trade-offs and systems they are particularly suited to.

Traditionally, non-empirical functionals have dominated this space, typically built on so-called semi-local approaches, which treat the XC functional in a localized fashion, with a local energy density depending on the local electronic density and its gradients. While exceedingly popular~\cite{VanNoorden2025TheseTime}, these approximations are limited by their lack of long-range interactions, introducing errors that limit the accuracy of DFT calculations. In response, many non-empirical methods leverage hybrid models, which include some fraction of Hartree exchange in combination with a semi-local DFA, but also incur substantial additional cost and still suffer from issues with accuracy in a variety of circumstances. In particular, non-empirical DFAs at all levels exhibit errors in band gap predictions~\cite{Jain2011ReliabilityGaps}, reaction barriers~\cite{Liu2025RevisitingKinetics}, and failure in systems with strong static correlation~\cite{Gibney2023UniversalCorrelation}. While there have been many post-hoc corrections proposed to address these challenges, including density correction~\cite{Vuckovic2019DensityDFT} and energetic adjustments such as dispersion correction~\cite{Goerigk2017ACorrection}, an alternative approach in the form of machine-learned (ML) DFAs has garnered much interest from the community in recent years~\cite{Li2021Kohn-ShamPhysics, Kasim2021LearningTheory, Kirkpatrick2021PushingProblem}. Such approximations promise to address the limitations of commonly used DFAs in an empirical fashion, often using differentiable physics codes to achieve these ends~\cite{Li2021Kohn-ShamPhysics, Kasim2022DQC:Chemistry, Zhang2022DifferentiableBeyond}.

One of the most interesting sets of systems on which non-empirical DFAs fail are strongly correlated geometries. These types of systems are typically poorly captured by DFT calculations or even higher-level methods such as coupled cluster (CC) theory~\cite{Bulik2015CanCorrelation, Sokolov2020QuantumEquivalents}, and have thus become a testbed for innovative DFA designs~\cite{Li2021Kohn-ShamPhysics, Kirkpatrick2021PushingProblem, Sokolov2026Quantum-enhancedFunctionals}. The simplest of these systems is the hydrogen dissociation curve, and a common component of all ML-DFAs which manage to correctly capture the energetics of this system is some form of non-local interaction. This non-locality enters either through Hartree exchange in some form of hybrid model~\cite{Kirkpatrick2021PushingProblem}, or through non-local design of the ML-DFA itself~\cite{Li2021Kohn-ShamPhysics, Sokolov2026Quantum-enhancedFunctionals}. Unfortunately, both of these approaches typically scale poorly, with reference~\cite{Kirkpatrick2021PushingProblem} in particular exhibiting $O(N^4)$ scaling. Li \textit{et al.}~\cite{Li2021Kohn-ShamPhysics} propose a global CNN in 1D to treat dissociated Hydrogen systems. However, it is not obvious how such an approach could reliably and efficiently be extended to a 3D system, as this global CNN scales $O(N^2)$. In 3D molecular systems with appropriate accuracy grids such as the standard grid 2 (SG-2)~\cite{Dasgupta2017StandardSG-3} even a single hydrogen nucleus requires evaluations on 7,094, rendering such scaling prohibitively expensive. Furthermore, the grid structure of a typical grid for molecules, composed from atom-centered components, follows the fuzzy construction by Becke~\cite{Becke1988AMolecules}, causing a highly irregular geometry with no obvious means of defining a convolution thereon. Sokolov \textit{et al.}~\cite{Sokolov2026Quantum-enhancedFunctionals} have suggested the usage of global quantum neural networks to circumvent these concerns, but benefits are only realized on a quantum computer, which limits applicability substantially.

Herein, we will construct a linearly sparse graph on the computational electronic grid for molecular systems by following the Exphormer procedure proposed by Shirzad \textit{et al.}~\cite{Shirzad2023Exphormer:Graphs}. We proceed by proposing a linearly scaling set of local graph edges which can be applied universally to computational grids of molecules, and enhancing this by a set of expander edges and a small number of globally connected reservoir nodes~\cite{Shirzad2023Exphormer:Graphs}. Expander constructions are particularly interesting for this type of application, as they allow for highly connected graphs with linearly scaling edge count~\cite{Hoory2006ExpanderApplications}. We use this construction to train a transformer DFA called Exphormer-XC on the full configuration interaction (FCI) Hydrogen dissociation curve, extending the differentiable quantum chemistry (DQC)~\cite{Kasim2022DQC:Chemistry} package to allow for Graph-NN DFAs. We use only densities, spin polarizations and spatial distances as inputs to the graph, and demonstrate that with the inclusion of all Exphormer components, the Hydrogen dissociation curve can be accurately captured. We furthermore show that different ablated versions of this estimator do not suffice to achieve the same result, with the expander contributions in particular proving key to achieve convergence of a Exphormer-XC. Finally, we extend our analysis to the planar $\mathrm{H_4}$ system near the square configuration, a notoriously challenging problem even for high-level coupled cluster (CC) calculations~\cite{Bulik2015CanCorrelation, Sokolov2020QuantumEquivalents}. We show that despite some issues with non-convergence due to the degeneracy in this system, Exphormer-XC is capable of capturing the correct energies of both the singlet and triplet state of this system.

\subsection{Related work}
Beyond the contributions mentioned above, much of the research into ML-DFAs has centered not on strongly correlated systems, but rather main group thermokinetics. There exist ML-DFAs focusing on semi-local approximations~\cite{Kasim2021LearningTheory, vonStrachwitz2026Data-efficientDFT}, and range limited approximations~\cite{Nagai2020CompletingMolecules, Schmidt2019MachineTheory, Zhuang2025MachineTheory, Lei2019DesignDescriptors, Dick2020MachineDensity}. Of these, only Schmidt \textit{et al.}~\cite{Schmidt2019MachineTheory} studied strongly correlated systems with mixed results. While they did achieve better energy errors on Hydrogen dissociation than non-empirical DFAs, errors remained well in excess of 1 kcal / mol and the calculations exhibited instabilities for longer interaction ranges. Another approach, proposed by Khan \textit{et al.}~\cite{Khan2025AdaptingLearning} includes a variable mixing fraction of Hartree exchange in a hybrid DFA, but has similarly not been applied to strongly correlated systems and requires relatively expensive Hartree exchange evaluations. Finally, molecular graph approaches have seen some interest in recent years, typically in replacing DFT calculations altogether in order to accelerate computation~\cite{Gong2023GeneralHamiltonian, Jrgensen2022EquivariantSolids}, but have also seen application to estimating the XC functional~\cite{Gao2024LearningFunctionals}. Once again, studies on strongly correlated systems have not been conducted, and these approaches differ conceptually from this work by taking nuclei as nodes in the graph, rather than computational grid points.

\section{Theory}
\subsection{Differentiable KS-DFT for ML-DFAs}
In Kohn-Sham density functional theory, the electrons of a system are collectively represented as a spatial density $n(\pos)$. This density is then found to be the one which minimizes the energy of the system $E[n]$, typically written as

\begin{equation}
    E[n] = T[n] + \int d\pos \ v_{\mathrm{ext}}(\pos)n(\pos) + E_H[n] + E_{XC}[n].
\end{equation}

\nin Here, $T[n]$, $v_{\mathrm{ext}}(\pos)$, and $E_H[n]$ are the well-established KS kinetic energy, external potential, and Hartree energy of the system. $E_{XC}[n]$ on the other hand represents the exchange-correlation energy, a collection of all energetic contributions not captured by the other terms. Typically, different approximations to this universal functional have been proposed, referred to as density functional approximations (DFAs). As the XC contributions remain a leading cause of error, particularly in strongly correlated systems such as a stretched hydrogen molecule (where many fail completely), much research has gone into both empirical and non-empirical DFAs. The most common construction follows a (semi-)local approach, with the overall energy written as

\begin{equation}
    E_{XC}[n] = \int d\pos \epsilon_{XC}(\gamma(\pos))n(\pos),
\end{equation}
\nin with $\gamma$ representing a set of parameters corresponding to the Taylor expansion of the spin-resolved density at different levels of approximation, namely
\begin{equation}
    \gamma(\pos) = \begin{cases}
        n_{\uparrow / \downarrow}(\pos) & \mathrm{Local\ density\ approximation\ (LDA)}\\
        n_{\uparrow / \downarrow}(\pos), \nabla n_{\uparrow / \downarrow}(\pos) & \mathrm{Generalized\ gradient\ approximation\ (GGA)}\\
        n_{\uparrow / \downarrow}(\pos), \nabla n_{\uparrow / \downarrow}(\pos), \nabla^2n_{\uparrow / \downarrow}(\pos) & \mathrm{meta-GGA}.
    \end{cases}
\end{equation}

ML-DFAs have been used to approximate either $E_{XC}$ directly or more commonly $\epsilon_{XC}$, the latter sometimes including other contributions than those outlined in $\gamma$. Both can be trained either using already converged densities $n$, or self-consistently within a differentiable KS solver. While the former is cheaper computationally~\cite{Kirkpatrick2021PushingProblem}, the latter has the advantage of ensuring that any XC functional obtained therein fulfills the self-consistency criterion, with energies, densities and potentials all arising from the same functional form and thus ensuring they are in agreement with one another. While both have seen successes, the latter has proven most promising for strongly correlated systems~\cite{Li2021Kohn-ShamPhysics, Sokolov2026Quantum-enhancedFunctionals}, and also given rise to a range of off-the-shelf differentiable KS solvers that can be used~\cite{Zhang2022DifferentiableBeyond, Kasim2022DQC:Chemistry, Herbst2021DFTK:Solids}. This latter approach will be used throughout our work, particularly by extending the DQC package~\cite{Kasim2022DQC:Chemistry} with \texttt{torch\_geometric}~\cite{Fey2025PyGGraphs} to admit graph based energy densities $\epsilon_{XC}$, which we shall now introduce briefly.
\subsection{Computational grids for molecular systems}
In practice, the XC energy is constructed on a quadrature grid, rather than continuous space, and in molecular systems, Becke's scheme~\cite{Becke1988AMolecules} is the default approach for such grid construction. Molecular grids here are constructed as the sum of contributions from overlapping, spherical, atom-centric grids. We may write the full term as
\begin{equation}
    E_{XC}[n] \approx \sum^{N_{\mathrm{atom}}}_a \sum_b^ {N_r(a)}w_b\sum_c^{N_{\Omega}(a, b)}w_{c}\epsilon_{XC}(\gamma(\pos_{abc}))n(\pos_{abc}),
\end{equation}

with each atomic contribution in turn being written as a spherical sum across $N_r(a)$ radial, and $N_{\Omega}(a, b)$ solid angle grid points. The weights $w_b$, $w_c$ are chosen to avoid singularities and discontinuities across the entire molecular grid and maintain consistent quadrature. The position of each grid point may be written as

\begin{equation}
    \pos_{abc} = \mathbf{R}_a + \begin{pmatrix}r_b \mathrm{sin}(\theta_c)\mathrm{cos}(\phi_c)\\r_b \mathrm{sin}(\theta_c)\mathrm{sin}(\phi_c)\\r_b \mathrm{cos}(\theta_c)\end{pmatrix}.
\end{equation}

\nin Here, $\mathbf{R}_a$ is the location of the $i$th nucleus, and $\theta_c$ and $\phi_c$ are sampled according to a Lebedev scheme, which ensures exact quadrature for functions up to a finite degree of spherical harmonic~\cite{Lebedev1976QuadraturesSphere}. The radial distances $r_b$ on the other hand may be sampled from a variety of different schemes, typically becoming exponentially denser towards the nucleus. Here, we will be using the double exponential scheme built into the SG-2 grid~\cite{Dasgupta2017StandardSG-3}. This grid can now be used to serve as the building blocks for the graph XC functional.
\subsection{Exphormer-XC}
In our work, we take the local XC density to depend not only on $\gamma(\pos_{abc})$, but to also include a contribution from an electronic graph $\mathcal{G}$. Following the Exphormer construction from Shirzad \textit{et al.}~\cite{Shirzad2023Exphormer:Graphs}, this graph may be written as

\begin{equation}
    \mathcal{G} = (\mathcal{V}_{\mathrm{grid}} \cup \mathcal{V}_{\mathrm{global}}, \mathcal{E}_{\mathrm{local}}\cup\mathcal{E}_{\mathrm{exp}}\cup\mathcal{E}_{\mathrm{global}}).
\end{equation}

\begin{figure}
    \centering
    \includegraphics[width=\linewidth]{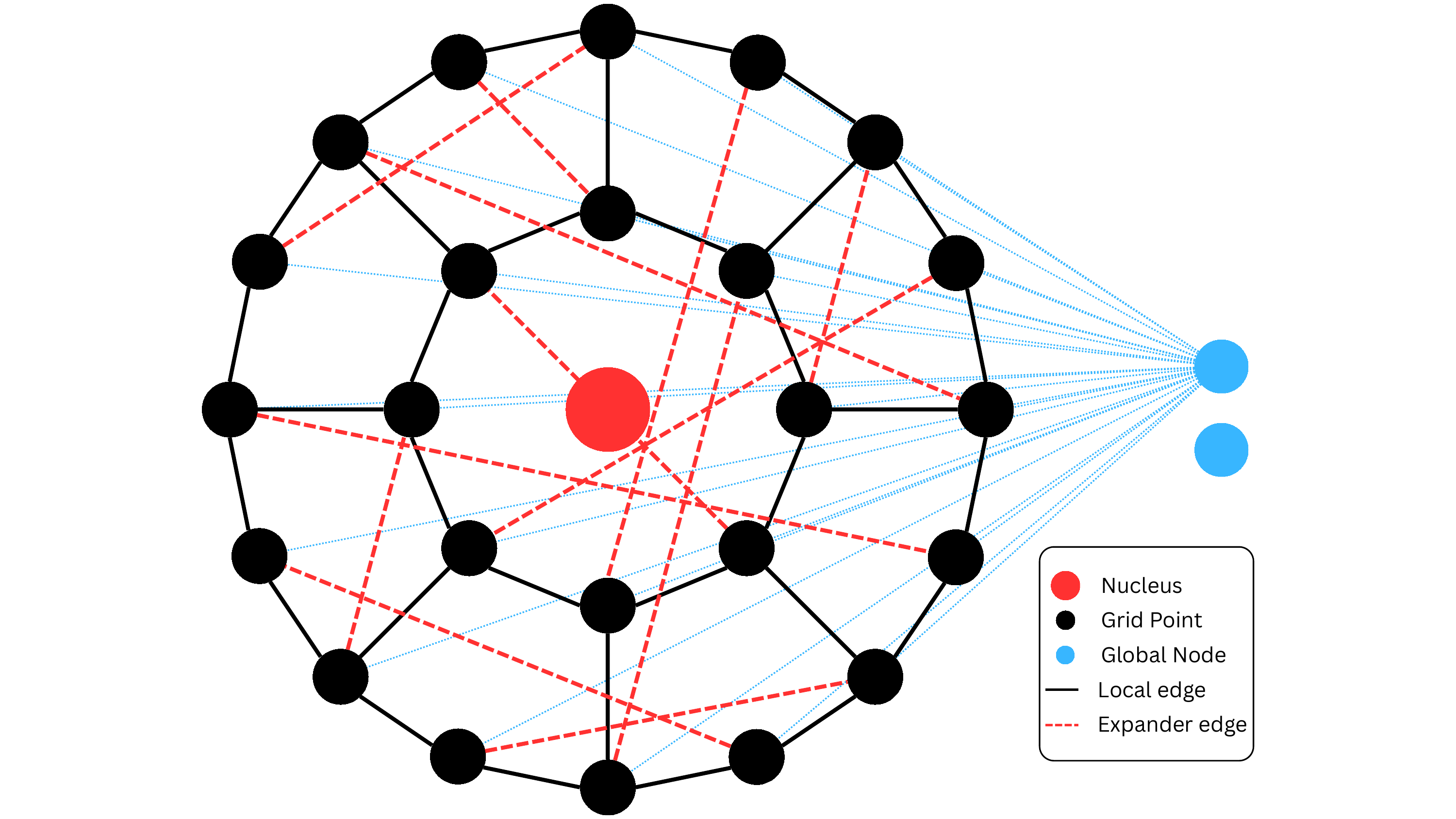}
    \caption{Diagram of the Exphormer-XC graph construction. Note that the expander edges have been reduced to degree 1 for visual clarity, and that edges are shown only for one global node for the same reason.}
    \label{fig:diagram}
\end{figure}

The three edge components are a local grid $\mathcal{E}_{\mathrm{local}}$, an expander $\mathcal{E}_{\mathrm{exp}}$, and a finite fixed number of globally connected ($\mathcal{E}_{\mathrm{global}}$) fictitious nodes $\mathcal{V}_{\mathrm{global}}$. We propose a straightforward, yet tunable construction for the first term, while following ref.~\cite{Shirzad2023Exphormer:Graphs} for the latter two. An illustration of this scheme is shown in figure~\ref{fig:diagram}. For local edges, we first connect nearest radial neighbours $\pos_{abc}$ and $\pos_{a(b+1)c}$ for $b=1,\hdots,N_r(a) -1$. Where adjacent shells share the same Lebedev degree, this is trivially possible, as they correspond to the same set of angular points scaled radially. Where this is not the case, the point in the inner shell is connected to its closest neighbour in the outer shell. For conciseness, we will simply write this contribution to the adjacency matrix between two nodes $\pos_{ijk}$ and $\pos_{abc}$ as

\begin{equation}
    A_{\mathrm{rad}} = \delta_{j(b+1)}\delta_{ia}\delta_{kc}.
\end{equation}

Next, we connect all nodes within a Lebedev shell that are sufficiently close to each other by Haversine distance, as determined by a variable cutoff $c$ such that

\begin{equation}
    A_{\mathrm{ang}} = \delta_{ia}\delta_{jb}\Phi(d_{kc} - d_{\mathrm{cutoff}}),
\end{equation}
\nin where $\Phi$ is the Heaviside function, $d_{kc}$ is the Haversine distance between two angular points and $d_{\mathrm{cutoff}}$ is defined as

\begin{equation}
    d_{\mathrm{cutoff}} = (1 + \alpha) \cdot \min_{l \neq k}d_{lk}.
\end{equation}
\nin Here, the minimization runs across all nodes in the same shell which yields a consistent contribution to the degree of each node, scaling slightly superlinearly with $\alpha$ until the shell saturates with edges, as shown in supplemental figure~\ref{fig:avg_degrees}. In addition to these local contributions, we add expander edges according to a simplified Friedman scheme~\cite{Shirzad2023Exphormer:Graphs}. We index the entire grid as vertices, and construct an expander with degree $d$ by concatenating $d/2$ copies of the grid vertex list, yielding

\begin{equation}
s = (\underbrace{1,\hdots,1}_{d/2},\underbrace{2,\hdots,2}_{d/2},\hdots,\underbrace{N_{\mathrm{grid}},\hdots,N_{\mathrm{grid}}}_{d/2}).
\end{equation}

\nin We then take a uniform random permutation $\pi$ on the set $\{1,\hdots,d N_{\mathrm{grid}}/2\}$ and construct the expander edges as

\begin{equation}
    \mathcal{E}_{\mathrm{expander}} = \{(s_i, s_{\pi(i)}), (s_i, s_{\pi^{-1}(i)}): 1\leq i \leq d N_{\mathrm{grid}} / 2\}.
\end{equation}

We validate this construction for some trial systems with the Ramanujan (spectral gap) expander criterion, as shown in supplemental figure ~\ref{fig:spectral_gap}. Next, we extend the set of $\mathcal{V}_{\mathrm{grid}}$ nodes by a small fixed number of nodes $K$ and connect them to all nodes in the base grid. These latter two steps mirror exactly the construction from Shirzad \textit{et al.}~\cite{Shirzad2023Exphormer:Graphs} on general graphs. Finally, we add a multi-layer, multi-head transformer as the neural network~\cite{Shi2020MaskedClassification}, where the per-head attention may be written in terms of feature vectors $\mathbf{x}_i$ of node $i$, and edge features $\mathbf{e}_{ij}$ for the edge between nodes $i$ and $j$ as
\begin{equation}
    f(\mathbf{x}_i, \mathcal{G}) =  W_1\mathbf{x}_i + \sum_{j\in \mathcal{N}(i)}\mathrm{softmax}\left(\frac{(W_2\mathbf{x}_i)^T(W_3\mathbf{x}_j + W_5\mathbf{e}_{ij})}{\sqrt {d}}\right)(W_4\mathbf{x}_j + W_5\mathbf{e}_{ij}).
\end{equation}

\nin Here, node features $\mathbf{x}_i$ at the input layer are simply the concatenated density $n = n_{\uparrow} + n_{\downarrow}$ and spin polarization $\zeta = (n_{\uparrow} - n_{\downarrow}) / n $ at node $i$, while edge features are Euclidean distance $r_{ij}$ between the nodes and an edge type categorical feature $t = 0, 1, 2$ to differentiate between local, expander, and global nodes. The full neural network with $K$ layers $f_k$ written as

\begin{equation}
F_{\mathrm{exp}}(\gamma; \mathcal{G}) = f_K \circ\hdots f_2\circ f_1(\gamma; \mathcal{G})
\end{equation}

is then taken as an enhancement factor of the local exchange-correlation energy density, which we may write as
\begin{equation}
    \tilde{\epsilon}_{XC}(\gamma; \mathcal{G}) = \epsilon_{XC}(\gamma) (1 + \beta F_{\mathrm{exp}}(\gamma; \mathcal{G})),
\end{equation}

\nin where $\beta$ is a learnable parameter initialized to zero, thus ensuring smooth transitions from the base DFA to the Exphormer-XC.

\section{Experiments on strongly correlated systems}

\subsection{Hydrogen dissociation curve}
\begin{figure}
    \centering
    \includegraphics[width=0.75\linewidth, clip, trim={500, 170, 500, 170}]{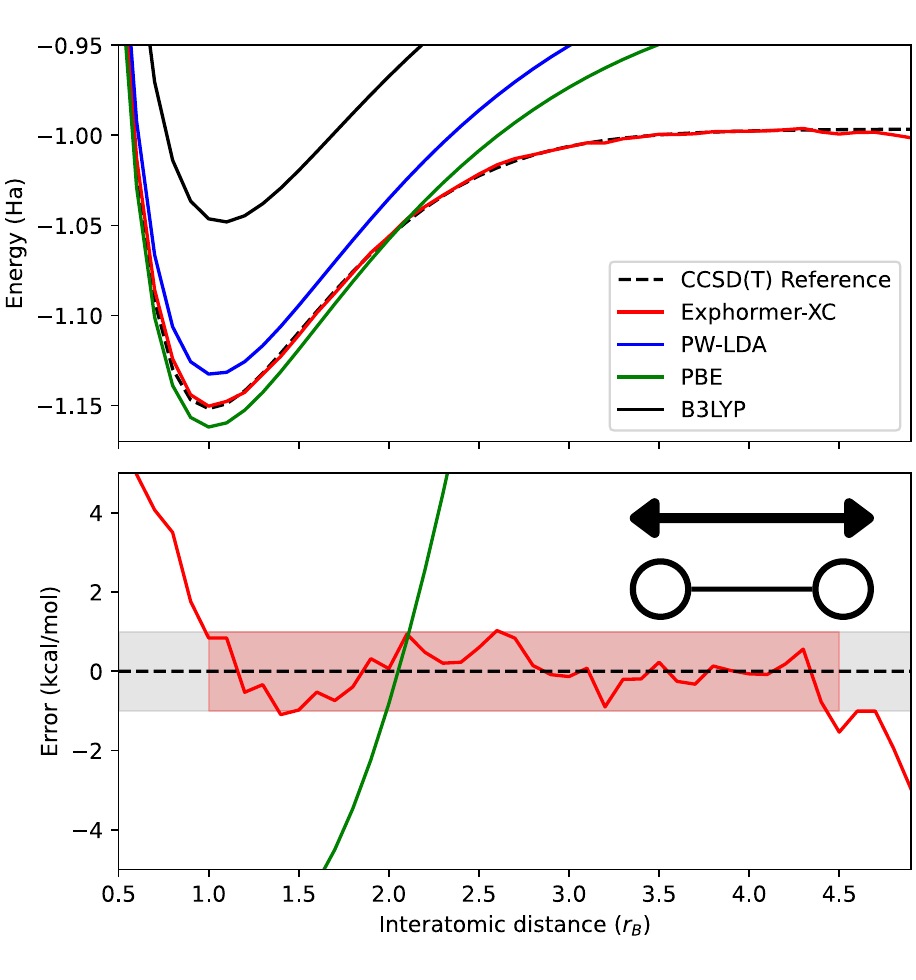}
    \caption{Dissociation curve of the Hydrogen molecule, with energies on top and errors with respect to FCI/6-31G at the bottom. The gray shaded area indicates 1 kcal / mol errors, with the red shaded error indicating the interpolative regime for training.}
    \label{fig:dissoc_curve}
\end{figure}

The Hydrogen dissociation curve is the most prototypical of all the strongly correlated systems under study. With all major non-empirical semi-local and hybrid XC functionals failing to capture its shape correctly, it is one of the systems that warrants close study with any novel DFA. In order to evaluate performance of Exphormer-XC on this system, we create a set of scaled geometries of the Hydrogen molecule parameterized by the scaling factor $S$. The interatomic distance is given by $R=S\cdot1.400r_{B}$, where $r_B$ is the Bohr radius and $1.400r_B$ the equilibrium distance of the Hydrogen molecule. The range of geometries varied from the strongly compressed regime ($S=0.5$) to the near-atomization limit ($S=5$). Total and atomization energies, as well as electronic densities for each geometry were calculated with PySCF~\cite{Sun2018PySCF:Framework} using full configuration interaction (FCI) in the 6-31G basis set. Originally, we trained Exphormer-XC on three training and one validation geometry akin to previous work~\cite{Li2021Kohn-ShamPhysics, Sokolov2026Quantum-enhancedFunctionals}, similarly relying on regularization from the differentiable KS solver. However, we found that the Exphormer architecture was too expressive with our chosen hyperparameters as shown in appendix~\ref{app:hyperparam}, resulting in overfitting on the small training and validation sets. This was mitigated by providing the model with more training and validation points along the curve ranging from $S=1$ to $S=4.5$, and led to test set errors less than 1 kcal / mol in the interpolative regime with respect to the reference data as shown in figure~\ref{fig:dissoc_curve}. This result is remarkable, as it has been shown conclusively that semi-local ML-DFAs are incapable of recovering this curve~\cite{Li2021Kohn-ShamPhysics}, and even modern non-empirical hybrid functionals designed explicitly to capture both strongly correlated and systems with delocalization error fail to capture this shape accurately~\cite{Khan2025SCANFunctional}. To further study how the dissociation limit is correctly captured with Exphormer-XC, we conduct an ablation study.

\subsection{Ablation study}
Due to the inability of semi-local DFAs, as well as many of their hybrid counterparts, to capture the Hydrogen dissociation curve at all, the ablation study was performed by simply investigating the training set performance of different ablated versions of the estimator. It has been well established that semi-local DFAs still struggle to fit the curve at all, even when machine-learned~\cite{Li2021Kohn-ShamPhysics}, so any achievement here will indicate performance specific to the estimator architecture chosen, rather than being indicative of any dataset constraints or other considerations. The details of each ablated model are explained in the supplemental material, but the summary of results are shown in table~\ref{tab:ablation_study}. Broadly, we explored an NN-LDA model first, serving as a purely local baseline. We then also compared a graph convolutional architecture~\cite{Kipf2016Semi-SupervisedNetworks} with and without distance edge scaling and an alternative NNConv architecture~\cite{Gilmer2017NeuralChemistry} which has been shown to perform well on molecular graphs and includes native edge embeddings rather than simple scaling. We then proceeded to consider a graph transformer without any expander contributions, before proceeding to the Exphormer both with / without distance edge embeddings, and with / without global nodes. We found that only the full Exphormer and the model without global nodes achieved 1 kcal / mol accuracy on the Hydrogen training set consisting of total and atomization energies for 6 different geometries along the dissociation curve. While both reached the desired threshold, the exclusion of global nodes substantially delayed the training convergence, taking an additional $21\%$ longer until the threshold was met. All other variations had detrimental effect on the training error, and were terminated due to early stopping when loss did not improve within 500 epochs or if their loss was caused by catastrophic convergence failures in the KS calculation, as was the case for the NNConv model. Overall, these results demonstrate that all components but the global nodes are strictly required for fitting the Hydrogen curve to sufficient accuracy within our framework, and that global nodes still present an advantageous contribution to the performance of the model. Next, we will investigate how this model performs on a more challenging system still, namely planar $\mathrm{H_4}$.

\begin{table}
    \centering
    \begin{tabular}{cccccc}
    \hline
        Model & Variant & Graph & Distance embedding & Epochs & MAE (kcal/mol)\\
        \hline\hline
        NN-LDA & - & \hasnt & \hasnt & 2,626 & 5.93\\
        Graph convolution & default & \has & \hasnt & 1,453 & 13.73\\
        Graph convolution & distance weighting & \has & \has & 1,452 & 13.97\\
        NNConv & - & \has & \has & 698 & 25,812.90\\
        TransformerConv & - & \has & \has & 1,214 & 2.82\\
        Exphormer & No distance embedding & \has & \hasnt & 754 & 4.58\\
        Exphormer & No globals & \has & \has & 780 & $\mathbf{<1}$\\
        Exphormer & Full exphormer & \has & \has & \textbf{641} & $\mathbf{<1}$\\
         \hline
    \end{tabular}
    \caption{Ablation study of different model components. Epochs given are until 1 kcal / mol is reached or until early stopping is triggered after 500 epochs of no improvement. Bold text indicates the best performing model(s).}
    \label{tab:ablation_study}
\end{table}

\subsection{Planar $H_4$}
Planar $\mathrm{H_4}$ in a near-square geometry exhibits not only strong correlation, but also a near-degeneracy in its spin states. The resulting quantum state is captured poorly both by DFT and CC methods, incorrectly predicting both the size and shape of the energy barrier posed by the square configuration. FCI calculations in the singlet ground state predict a parabolic energy barrier with relatively small amplitudes while DFT and CC methods typically predict energetic cusps of substantially larger amplitude, with some CC methods furthermore incorrectly predicting a minimum rather than a maximum~\cite{Bulik2015CanCorrelation}. If the degeneracy is not broken explicitly by enforcing the symmetry of the system, even the FCI method can incorrectly capture the excited triplet state in this system. The combination of these effects make this system an exceedingly challenging one to model with any quantum chemistry method.

\begin{figure}
    \centering
    \includegraphics[width=\linewidth, clip, trim={400, 200, 400, 170}]{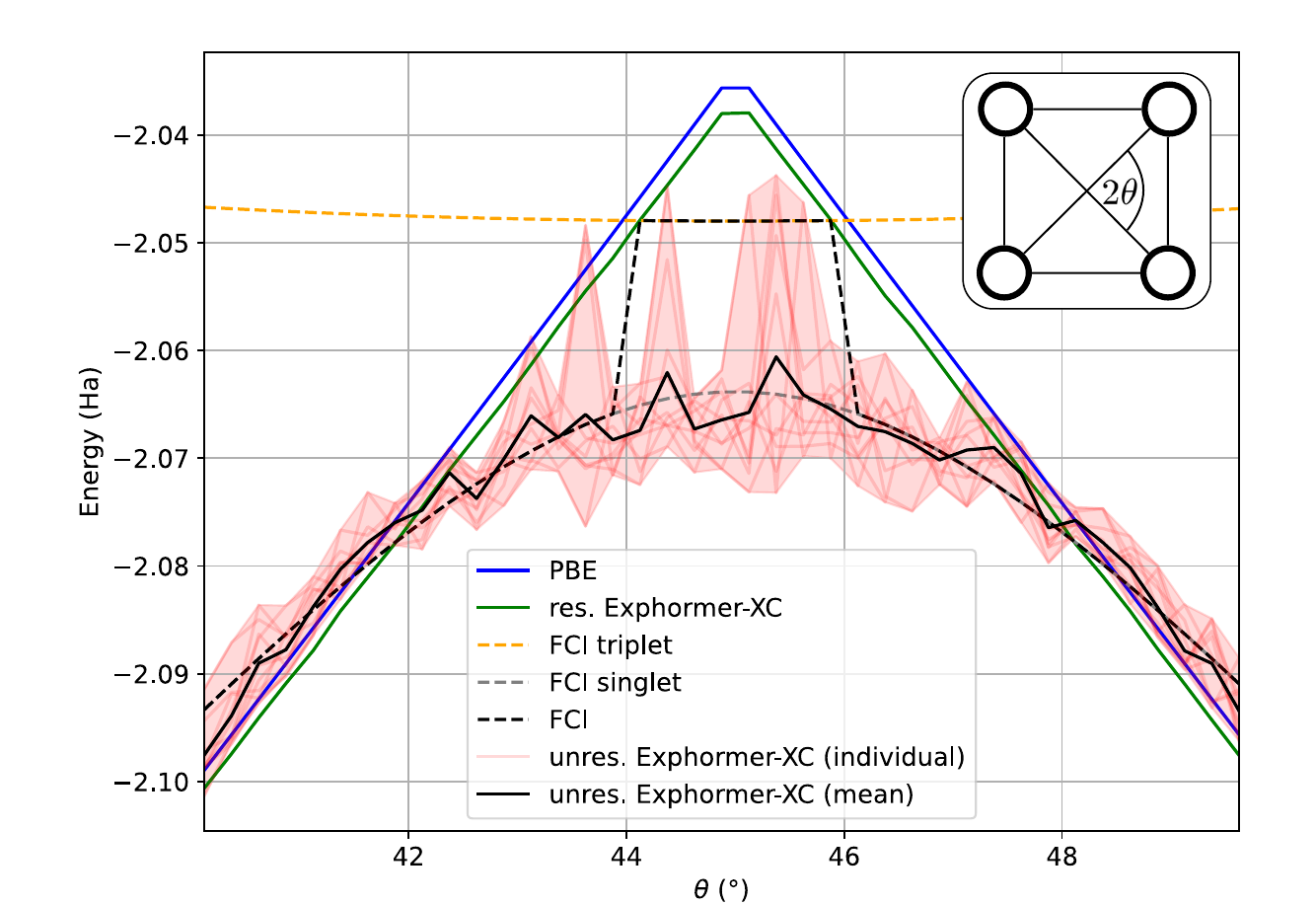}
    \caption{The energy barrier in planar $\mathrm{H_4}$ predicted by different quantum chemistry methods. The black dashed line indicates FCI without explicit symmetry breaking. Red shaded area is the envelope of all unrestricted calculations, with faint red lines corresponding to individual predictions and the solid black line indicating the mean.}
    \label{fig:h4_curve}
\end{figure}

In order to study whether Exphormer-XC is capable of addressing this problem, we varied the angle of the geometry around the energy barrier formed by the square configuration, with nuclei located at $\pm R\sin(\theta)$ and $\pm R \cos(\theta)$. The relevant regime was parametrized as $\theta=[40\degree, 50\degree]$ with $R=2.0r_B$. Here, the near-degeneracy gives rise to a predicted spin state transition in the FCI, from the singlet state for $\theta < 44\degree$ and $\theta > 46\degree$ to the higher energy triplet state with $44\degree \leq \theta \leq 46\degree$. The calculation in PySCF can be stabilized in the singlet state by explicitly including symmetry of the system thus breaking the spin degeneracy and capturing the actual ground state. The failure of typical DFAs to correctly capture the behaviour of this system is illustrated by the semi-local PBE functional~\cite{Perdew1996GeneralizedSimple} as shown in figure~\ref{fig:h4_curve} which differs qualitatively from the FCI reference in predicting a sharp cusp, where the FCI ground state predicts a parabolic energy barrier, and quantitatively in overpredicting the energy barrier by $0.03$ Ha, or $18.8$ kcal / mol. The local PW-DFA~\cite{Perdew1992AccurateEnergy} and hybrid B3LYP model~\cite{Becke1993ATheories}, were found to exhibit similar but quantitatively worse behaviour, with only a constant energy shift separating the methods, and were thus omitted for clarity.

Exphormer-XC in a spin-restricted ($n_{\uparrow} = n_{\downarrow}$) KS-calculation was found to produce much the same behaviour as the PBE-DFA irrespective of base model, with only a negligible shift in energy and no qualitative differences. By contrast, the unrestricted Exphormer with a PBE base model was capable of much more closely fitting the FCI energy, but exhibited convergence issues in the KS calculations. While non-convergence is a relatively generic issue, the level of non-convergence can differ substantially, and it can have different causes. Past work with differentiable KS solvers and ML-DFAs found that improper initialization can give rise to catastrophic non-convergence of calculations, for example~\cite{vonStrachwitz2026Data-efficientDFT}. Another cause, however, are (near-)degeneracies of the system. As excited states constitute saddle points in the KS solution, their presence can induce the failure of the DFT calculation to converge adequately. Specifically, the calculation may jump back and forth between states during the self-consistency cycles, impeding convergence to one or the other. This is precisely the effect observed for the $\mathrm{H_4}$ systems here, which we illustrate by plotting 10 separate but identical calculations across the entire energy barrier, thus capturing the stochastic error induced by non-convergence. On the whole, the energies predicted still agree more closely to the FCI calculations than the spin-restricted Exphormer-XC or other DFAs. Calculations within $\pm 1.5\degree$ of the exact square configuration in particular occasionally exhibit energies consistent with the triplet state, as well as those consistent with the singlet state, indicating that the model is capable of capturing the energies of both states while only being trained on one. Furthermore, the amplitude of the residual in the self-consistent calculation decreases further from the cusp, and in all cases, the energetic prediction is still closer to the FCI singlet state than any of the other DFAs, with the mean across calculations following the state relatively closely.

While the convergence errors here limit the strength of the conclusions one can draw, these results nevertheless seem to indicate that the model could converge properly if the degeneracy was broken by including symmetry of the system explicitly, as for the FCI calculations. Unfortunately, the differentiable solver used in our study does not have this capacity, and due to the novel computational ansatz presented here no other quantum chemistry codes are capable of this representation. Thus, this hypothesis cannot be tested here.

\section{Limitations}

Although generalization of Exphormer-XC to other chemical systems has not been studied here, it is expected to perform well on those systems given sufficient training data, as models trained within the same code have been shown to faithfully reproduce ground truth data~\cite{vonStrachwitz2026Data-efficientDFT}. The main limitation is likely the amount of data required, given that this model is substantially more expressive than semi-local approximations.

The choice of the electronic grid as the base structure on which to define its graphs is expected to be the main limitation of Exphormer-XC. Rather than a smaller, molecular graph approach, an electronic graph leverages a much larger number of nodes, and is not uniquely defined, but rather dependent on the chosen spatial grid. This, in principle, allows it to leverage much richer spatial information, but it does not naturally transfer to other grids, potentially requiring retraining for different resolutions. This limitation is inherent to this kind of approach and mitigated primarily by the favourable scaling of the graph architecture.

In addition, the novelty of our computational architecture would require substantial amounts of additional work to make our model compatible with other codes, limiting its usage unless compatibility with graphs on computational grids is implemented more broadly.

Furthermore, the non-convergence observed in the $\mathrm{H_4}$ case constitutes another obvious limitation of the application to that trial system, but this could potentially be remedied by including symmetry explicitly in the KS calculation as discussed above.

Finally, application to the $\mathrm{H_2^+}$ system has not been studied here, although it is a prototypical system for studies which consider strongly correlated systems. This omission was made as previous work~\cite{Li2021Kohn-ShamPhysics, Sokolov2020QuantumEquivalents} rely on a "self-interaction gate" to correctly capture the energetics of this system. This gate simply reverts to Hartree exchange in the case of a single electron system (i.e. $\mathrm{H_2^+}$) and thus no learning occurs at all for this system. Such a gate could trivially be included to make Exphormer-XC viable on that system, and was thus omitted.

\section{Conclusion}
In this paper, we have introduced the Exphormer-XC graph DFA as the first linearly scaling ML-DFA capable of capturing the hydrogen dissociation curve. Building on previous applications of ML-DFAs to strongly correlated systems, we provide a careful construction that balances the non-locality required to capture this regime with computational cost, substantially improving on the previous best quadratic scaling in the field. We furthermore demonstrate that this construction is necessary to achieve accurate modeling of non-local effects, and that a simpler model would not suffice. Extending this success to a more challenging setting, we explored the application of our model to planar $\mathrm{H_4}$. Here, Exphormer-XC shows strong potential for accurately capturing the energetics of this system, a task which has only been achieved by a quantum computer DFA in the past.
While the application to main group thermochemistry tasks remains an open direction for future work, the strong performance of our model on strongly correlated systems is indicative of a highly capable DFA. Taken together, these results open high-accuracy non-local ML-DFAs to a much broader range of applications, many of which have remained out of reach due to the poor scaling of previous approaches.

\bibliographystyle{unsrt}
\bibliography{references}

\newpage
%\input{checklist.tex}
%\newpage
\appendix

\begin{figure}
    \centering
    \includegraphics[width=\linewidth]{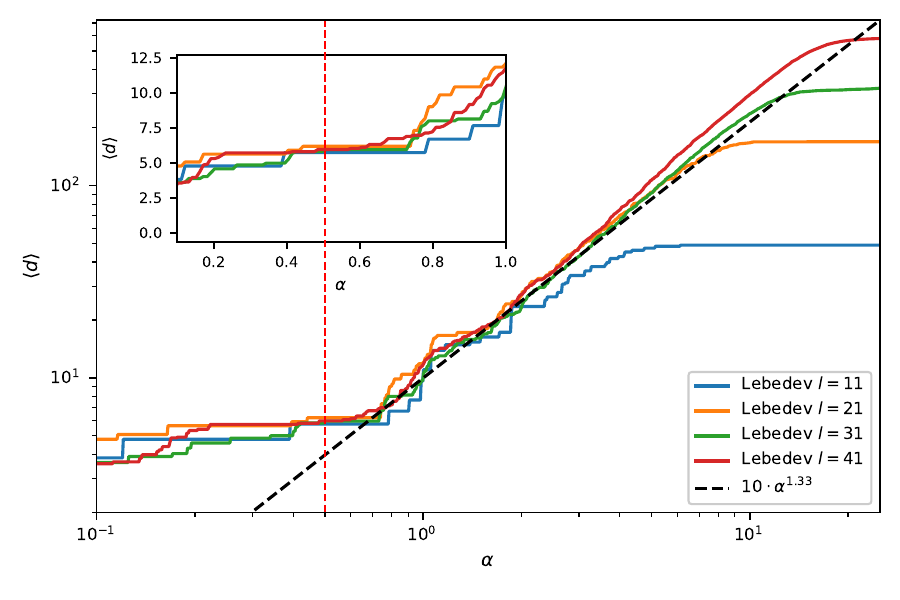}
    \caption{Average degree against $\alpha$ parameter for varying Lebedev orders. The dashed vertical line indicates the value of $\alpha$ used throughout this work.}
    \label{fig:avg_degrees}
\end{figure}
\section{Graph construction}
\subsection{Local degree}
Average degree of the local graph $2|\mathcal{E}_{\mathrm{local}}| / |\mathcal{V}_{\mathrm{grid}}|$ remains constant with system size and resolution, as each component introduces a fixed number of edges per vertex. The only component for which scaling with resolution is not immediately obvious from the construction is the angular component, governed by the parameter $\alpha$. As the grid is constructed from multiple Lebedev shells in each nucleus, the scaling can be seen by comparing different Lebedev orders at the same value $\alpha$, as shown in figure~\ref{fig:avg_degrees}. For the value of $\alpha=0.5$ used throughout our work, the average degree is identical for a range of Lebedev orders, thus guaranteeing linear scaling within that range of orders. For larger $\alpha$ values, the connectivity saturates for lower Lebedev orders, as all points within a shell are connected. In this regime, the scaling with resolution will be superlinear, but scaling with system size will remain linear.

\begin{figure}
    \centering
    \includegraphics[width=\linewidth]{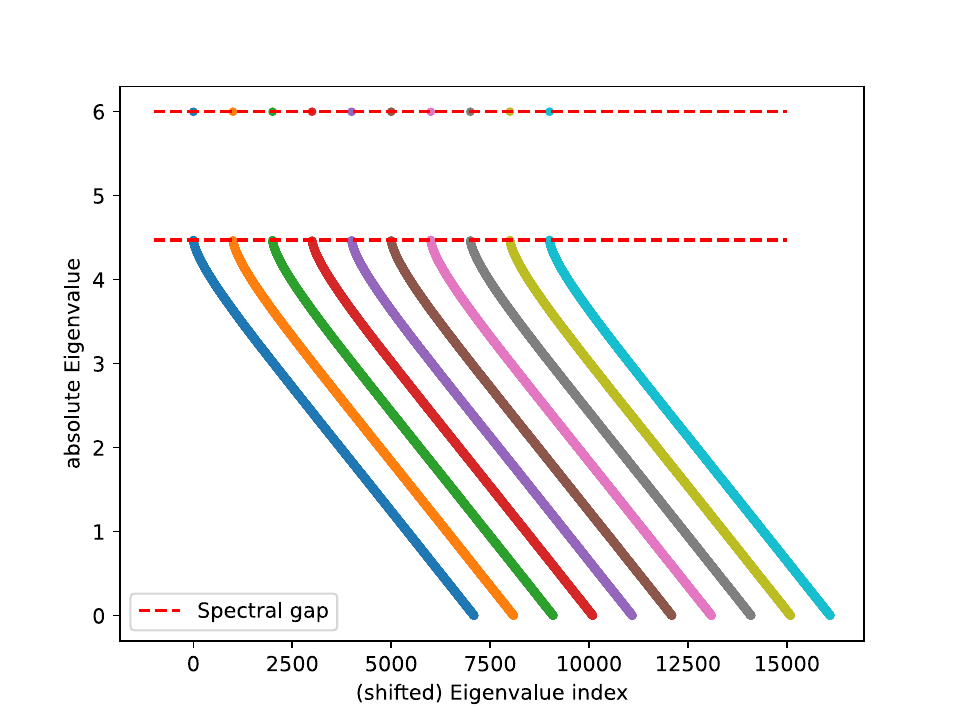}
    \caption{The spectrum of 10 instances of a Friedman graph construction scheme for a d-regular graph with 7,094 vertices and $d=6$, corresponding to the SG-2 grid for a single Hydrogen atom.}
    \label{fig:spectral_gap}
\end{figure}

\subsection{Expander validation}
The simplified Friedman construction outlined in the main text is validated by the Ramanujan criterion on the spectral gap, which is given for eigenvalues $\lambda_i$ of the adjacency matrix $A$ of a d-regular graph with $\lambda_1 > \lambda_2 > \hdots > \lambda_{|\mathcal{V}_{\mathrm{grid}}|}$ as

\begin{equation}
    |\lambda_1| - \max\limits_{i >1}|\lambda_i| \geq 2\sqrt{d-1}.
\end{equation}

While validation of all systems where Exphormer-XC may be applied is too expensive to perform regularly, the Friedman expander construction guarantees~\cite{Friedman2003RelativeGraphs} a spectral gap of $2\sqrt{d-1} - \varepsilon$, for any $\varepsilon > 0$ with probability $1 - O(|\mathcal{V}_{\mathrm{grid}}|^{-\xi})$ and

\begin{equation}
    \xi = \left|\frac{\sqrt{d -1} + 1}{2}\right|.
\end{equation}

Thus, failures are most likely for the smallest values of $d$ and $|\mathcal{V}_{\mathrm{grid}}|$, which are embodied by a single Hydrogen atom in the SG-2 basis in our construction. We therefore study the spectral gap for ten-fold random expander construction on this system, and found that the spectral gap criterion was met for all 10 instances of the expander as shown in figure~\ref{fig:spectral_gap}. For the smallest system actually studied in our work, namely $\mathrm{H_2}$, the probability of not constructing a Ramanujan graph with the same procedure for $d=6$ is $2^{-(\sqrt{5}+1)/2}\approx0.33$ times that of the single Hydrogen atom. Thus, the construction is assumed to yield a Ramanujan graph throughout this work.

\section{Training details}
\subsection{Optimizer and loss function}
Training was conducted using DQC~\cite{Kasim2022DQC:Chemistry}, with the ADAM optimizer~\cite{Kingma2014Adam:Optimization} and a learning rate of $5\cdot10^{-4}$. Weight decay of $5\cdot10^{-5}$ was employed. Mean absolute error was used across all objectives, which were total energy loss, atomization energy loss and localized energy loss (LEL). While the former two are standard error terms, the localized energy loss follows the construction by von Strachwitz \textit{et al.}~\cite{vonStrachwitz2026Data-efficientDFT} and is not to be confused with the LES defined in the literature~\cite{Polak2024Real-spaceFunctionals}, which is used to similar effect but with distinct motivation and derivation.

\subsection{Hyperparameters}\label{app:hyperparam}
The hyperparameters used for Exphormer-XC in the case of the Hydrogen dissociation curve are given in table~\ref{tab:hyperparameters}. The exact same set of hyperparameters was used for the $\mathrm{H_4}$ study, with the exception of the base model, which was instead set to PBE~\cite{Perdew1996GeneralizedSimple}.
\begin{table}[h!]
    \centering
    \begin{tabular}{cc}
    \hline
        Parameter & Value\\
         \hline\hline
         Graph parameters & \\
         \hline
        $\alpha$ & 0.5\\
        d & 6\\
        K & 10\\
         \hline
        Transformer parameters & \\
        \hline
        \# heads & 3\\
        \# layers & 4\\
        \# channels & 32\\
        \hline
        $\epsilon_{XC}$ & PW-LDA~\cite{Perdew1992AccurateEnergy}\\
         \hline
    \end{tabular}
    \caption{Hyperparameters of the Exphormer-XC used for $\mathrm{H_2}$ dissociation}
    \label{tab:hyperparameters}
\end{table}
\subsection{Further computational details}
All computations were conducted on an Intel(R) Xeon(R) w5-2465X desktop CPU with individual calculations taking seconds to minutes, depending on convergence of the calculation. Training of all models was continued to convergence or early stopping, and took less than 24 hrs for all models presented both in the main experiments and the ablation study. Initial setup times for calculations were driven primarily by the expander construction, as angular distances for local graph construction were saved alongside the Lebedev quadrature grids to avoid repeat calculation. Once an expander was constructed, it was cached and reused during training, but not during inference, where new expanders were generated for each system.
\section{Ablation study models}
The models used in the ablation study are listed below with relevant references and technical details specific to each construction.
\subsection{NN-LDA}
The simplest model considered is a purely local NN-LDA with 4 layers, 32 nodes per layer and softplus nonlinearities, following the construction by von Strachwitz \textit{et al.}~\cite{vonStrachwitz2026Data-efficientDFT}.
\subsection{Graph convolution}
The graph convolutional neural network by Kipf \& Welling~\cite{Kipf2016Semi-SupervisedNetworks}, with layers constructed as

\begin{equation}
    \mathbf{x}_i' = \sum_{j \in \mathcal{N}(i) \cup\{i\}}\frac{1}{\sqrt{d(i)d(j)}}\left(W\mathbf{x}_j\right) + \mathbf{b}.
\end{equation}

Here, $W$ and $\mathbf{b}$ are learnable weights and biases, and $d$ refers to the degree of each node. Four layers with 32 nodes each are used, combined with ReLU nonlinearities and optional edge weighting by $1 / \sqrt{r_{ij}}$.
\subsection{NNConv}
The NNConv model proposed for neural message passing in quantum chemistry~\cite{Gilmer2017NeuralChemistry}. Here, edge features are explicitly included as the inputs to the model, yielding the contributions from neighbouring vertices as

\begin{equation}
        s_{\mathcal{N}}(i) = \sum_{j\in N(i)}\mathbf{x}_j \cdot g_{\theta}(\mathbf{e}_{ij}),
\end{equation}

with a feed-forward neural network construction $g_\theta$, containing once again 4 layers with 32 nodes and ReLU nonlinearities.
\subsection{TransformerConv}
The Exphormer-XC but with all non-local contributions removed. The graph is given simply by $\mathcal{G}=(\mathcal{V}_{\mathrm{grid}}, \mathcal{E}_{\mathrm{local}})$.
\subsection{Exphormer without distance embedding}
The Exphormer-XC as presented in the main text, but with distances $r_{ij}$ excluded from the edge features, which instead are simply reduced to $\mathbf{e}_{ij}=(t)$.
\subsection{Exphormer without global nodes}
The Exphormer-XC as presented in the main text, but with $\mathcal{V}_{\mathrm{global}} = \mathcal{E}_{\mathrm{global}} = \varnothing$.
\subsection{Full exphormer}
The Exphormer-XC construction presented in the main text with all components included.
\end{document}